\documentclass[aps,twocolumn,superscriptaddress]{revtex4-1}

\usepackage{xspace, microtype}
\usepackage{amsmath, amssymb, stmaryrd, bm}
\usepackage{graphicx}
\usepackage{booktabs}
\usepackage{amsthm}

\newenvironment{rcases}
{\left.\begin{aligned}}
{\end{aligned}\right\rbrace}

\newcommand{\includetikz}[2]{%
  \includegraphics{#2.pdf}%
}

\newcommand\bX{\ensuremath{\bm X}\xspace}
\newcommand\bY{\ensuremath{\bm Y}\xspace}
\newcommand\PI[2]{\ensuremath{I_{\partial}^{#1\rightarrow #2}}\xspace}
\newcommand\IR[2]{\ensuremath{I_{\cap}^{#1\rightarrow #2}}\xspace}
\newcommand\wmsphi{\ensuremath{\Phi^{\mathrm{WMS}}}\xspace}
\newcommand{\join}{\ensuremath{\vee\xspace}}
\newcommand{\meet}{\ensuremath{\wedge\xspace}}
\newcommand{\phiid}{\ensuremath{\Phi\mathrm{ID}}\xspace}

\newtheorem{theorem}{Theorem}%

\newtheorem{proposition}[theorem]{Proposition}

\newtheorem{definition}{Definition}

\begin{document}

\title{Beyond integrated information: A taxonomy of information dynamics phenomena}

\author{Pedro A.M. Mediano}
\affiliation{Department of Computing, Imperial College London, London SW7 2RH}
\thanks{P.M. and F.R. contributed equally to this work.\\E-mail: \{pmediano, f.rosas\}@imperial.ac.uk}

\author{Fernando Rosas} 
\affiliation{Center for Psychedelic Research, Department of Medicine, Imperial College London, London SW7 2DD}
\affiliation{Data Science Institute, Imperial College London, London SW7 2AZ}
\affiliation{Center for Complexity Science and Department of Mathematics, Imperial College London, London SW7 2AZ}
\thanks{P.M. and F.R. contributed equally to this work.\\E-mail: \{pmediano, f.rosas\}@imperial.ac.uk}

\author{Robin L. Carhart-Harris} 
\affiliation{Center for Psychedelic Research, Department of Medicine, Imperial College London, London SW7 2DD}

\author{Anil K. Seth}
\affiliation{Sackler Center for Consciousness Science, Department of Informatics, University of Sussex, Brighton BN1 9RH}

\author{Adam B. Barrett}
\affiliation{Sackler Center for Consciousness Science, Department of Informatics, University of Sussex, Brighton BN1 9RH}
\affiliation{The Data Intensive Science Centre, Department of Physics and Astronomy, University of Sussex, Brighton BN1 9QH, UK}

\date{\today}

\begin{abstract}

Most information dynamics and statistical causal analysis frameworks rely on
the common intuition that causal interactions are intrinsically pairwise --
every `cause' variable has an associated `effect' variable, so that a `causal
arrow' can be drawn between them. However, analyses that depict
interdependencies as directed graphs fail to discriminate the rich variety of
modes of information flow that can coexist within a system. This, in turn,
creates problems with attempts to operationalise the concepts of `dynamical
complexity' or `integrated information.' To address this shortcoming, we
combine concepts of partial information decomposition and integrated
information, and obtain what we call \textit{Integrated Information
Decomposition}, or $\Phi$ID. We show how $\Phi$ID paves the way for more
detailed analyses of interdependencies in multivariate time series, and sheds
light on collective modes of information dynamics that have not been reported
before. Additionally, $\Phi$ID reveals that what is typically referred to as
`integration' is actually an aggregate of several heterogeneous phenomena.
Furthermore, $\Phi$ID can be used to formulate new, tailored measures of
integrated information, as well as to understand and alleviate the limitations
of existing measures.

\end{abstract}

\maketitle

How can we best characterise the plethora of dynamical phenomena that can
emerge in a system of interdependent components? Progress on this question will
enable important advances in our understanding, engineering and control of
multivariate complex systems, including the human
brain~\cite{kelso1995dynamic}, the global climate~\cite{runge2019inferring},
macroeconomics \cite{Dosi:2019}, and many more. A popular approach to study
such systems is to portray their interdependencies as a directed graph of
non-mediated dependencies from past to future events (for example with Granger
causality~\cite{bressler2011wiener}); and then to analyse this graph. However,
this approach has a serious limitation that is rarely acknowledged: it only
considers statistical causation acting from single `cause' variables to single
`effect' variables (or sets of variables), thus neglecting possible
higher-order causal interactions.

The above limitation is rooted in the misleading intuition that information
dynamics can be reduced to \textit{storage} and \textit{transfer} phenomena.
Accordingly, a number of theoretical frameworks have tried to assess the
dynamical complexity -- understood as the amount of information transfer -- of
various systems using one-dimensional metrics. A remarkable example of this is
found in the neuroscience literature, where it has been proposed that a key
feature of the neural dynamics underpinning advanced cognition, flexible
behaviour, and ultimately consciousness, can be captured by a single number
that accounts for the ability of the system to `integrate information.' There
have been several operationalisations of this notion, including the various
$\Phi$ measures in Integrated Information Theory
(IIT)~\cite{Tononi1994,Balduzzi2008,Oizumi2014} and Causal Density
(CD)~\cite{Seth2011}; however, these measures have been shown to behave
inconsistently~\cite{van2003neural,Mediano2019,Rosas2019}, making empirical
applications difficult to interpret. Other attempts to explain information
dynamics solely in terms of storage and transfer have been shown to be
similarly unsuccessful~\cite{james2016information}.

As a possible way forward, Lizier \cite{Lizier2010} postulated a third
category, \textit{information modification}, which informally appeals to the
notion of `computation,' although it still remains both theoretically and
practically imprecisely defined. Following Lizier's insight, we pursue a
multi-dimensional description of dynamical complexity, which can disentangle
qualitatively different modes of information dynamics and statistical
causality. Our approach is based on the partial information decomposition (PID)
framework~\cite{Williams2010}, which breaks down the information that multiple
source variables carry about a (single) target variable into redundant, unique
and synergistic components. Applying PID to a stochastic dynamical system
setting, we consider the decomposition of the whole set of `cause'- and
`effect'-type informational relationships, and obtain what we call the
Integrated Information Decomposition, $\Phi$ID. This new framework sheds light
on modes of information dynamics that have not been previously reported, and
which most statistical causation frameworks ignore. Additionally, $\Phi$ID
allows us to show how existing one-dimensional measures of integrated
information conflate qualitatively different phenomena.

\section*{Decomposing multivariate information}

Consider two interdependent processes that are measured at regular time
intervals. The \textit{excess entropy}~\cite{crutchfield2003regularities} of
these processes, $\mathbf{E}$, is the total amount of (Shannon) information
that is transferred through these processes from past to future, which is a
well-known metric to assess dynamical complexity~\cite{grassberger1986toward}.
While $\mathbf{E}$ is in general hard to compute~\cite{crutchfield2009time},
for Markovian processes it simplifies to
\begin{equation}\label{eq:excess_entropy_markov}
\mathbf{E} =  I(X_1,X_2; \; Y_1,Y_2)~,
\end{equation}
where $\bX$ and $\bY$ denote the states at times $t$ and $t+1$ respectively,
and the subscript denotes variable index. We consider the decomposition of
$\mathbf{E}$ into modes of information dynamics, focusing on systems with
Markovian dynamics, leaving extensions to processes with memory for future
work.

\subsection*{Forward and backward information decomposition}

Our approach is to decompose $\mathbf{E}$ using the principles of the
\textit{Partial Information Decomposition} (PID) framework \cite{Williams2010}.
By focusing on how information flows from past to future, one can consider a
\textit{forward PID} that decomposes the information provided by $X_1$ and
$X_2$ about the joint future $(Y_1,Y_2)$ as
\begin{align}\label{eq:forward}
\mathbf{E} =&\:\texttt{Red}(X_1,X_2;Y_1Y_2 ) + \texttt{Un}(X_1;Y_1Y_2|X_2) \nonumber\\
&+ \texttt{Un}(X_2;Y_1Y_2|X_1) + \texttt{Syn}(X_1,X_2;Y_1Y_2). \nonumber
\end{align}
Intuitively, $\texttt{Red}(X_1,X_2;Y_1Y_2 )$ corresponds to \textit{redundant
information} provided by both $X_1$ and $X_2$ about $Y_1Y_2$~\footnote{We use $Y_1Y_2$ as a shorthand notation for the random vector $(Y_1,Y_2)$.};
$\texttt{Un}(X_1;Y_1Y_2|X_2)$ (resp. $\texttt{Un}(X_2;Y_1Y_2|X_1)$) refers to 
the \textit{unique information} that only $X_1$ (resp. $X_2$) provides about
$Y_1Y_2$; and finally, $\texttt{Syn}(X_1,X_2;Y_1Y_2)$ accounts for the information that
$X_1$ and $X_2$ provide about $Y_1Y_2$ only when they are observed together,
henceforth called \textit{synergistic information}~\footnote{$\texttt{Syn}(X_1,X_2;Y_1Y_2)$ was in fact proposed in Ref. \cite{Griffith2014} as a measure of integrated information.}.

An equivalent decomposition can be built by considering the information that
$Y_1$ and $Y_2$ contain about the past state $(X_1,X_2)$. Correspondingly, a
\emph{backward PID} is given by
\begin{align}%
\mathbf{E} =&\: \texttt{Red}(Y_1,Y_2; X_1X_2 ) + \texttt{Un}(Y_1; X_1X_2|Y_2) \nonumber\\
&+ \texttt{Un}(Y_2; X_1X_2|Y_1) + \texttt{Syn}(Y_1,Y_2; X_1X_2 ). \nonumber
\end{align}
The forward and backward PID are related to the notions of \emph{cause} (forward) and \emph{effect} (backward) information in IIT. 
These two information decompositions provide complementary, but overlapping descriptions of the system's dynamics. 
The next section explains how they can be unified in a single and encompassing description.

\section*{Integrated information decomposition: $\Phi$ID}

This section develops the mathematical framework of our contribution. The goal is to provide a decomposition of $\mathbf{E}$ similar to the two above, but that applies to both cause and effect information simultaneously. To do this, we solve PID's limitation of having only one single target variable and move towards %
multi-target information decompositions.

\subsection*{Double-redundancy lattice}

Let us begin by considering the redundancy lattice \cite{Williams2010}, which
is used in PID to formalise our intuitive understanding of redundancy.
Let $\mathcal{A}$ be the collection given by
\begin{equation}
\mathcal{A} := \{ \{1\}, \{2\}, \{1,2\}, \{ \{1\},\{2\} \} \}, 
\end{equation}
which are all the sets of subsets of $\{1,2\}$ where no element is contained in
another~\footnote{In a general $N$-variable case, $\mathcal{A}$ is the set of
antichains on the lattice $(\mathcal{P}(\{1, ... , N\}), \subseteq)$, discussed
in \cite{Williams2010}. We focus on the bivariate case for clarity, although
our results hold for any $N$.}.

\begin{figure}[ht]
  \centering
  \includetikz{tikz/}{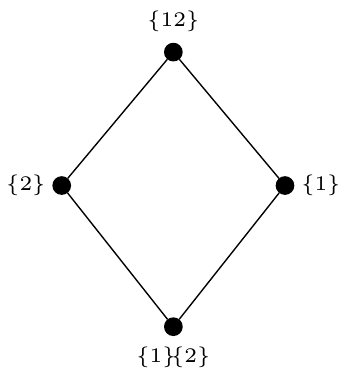}
  \caption{%
  Lattice of %
  nodes in $\mathcal{A}$ arranged according to the partial ordering $\preceq$.}
  \label{fig:single_lat}
\end{figure}

The elements of $\mathcal{A}$ have a natural (partial) order relationship: for $\bm
\alpha, \bm \beta \in \mathcal{A}$, one says that $\bm \alpha \preceq \bm \beta$
if for all $b\in\bm \beta$ there exists $a\in\bm\alpha$ such that $a \subset
b$ \cite{Williams2010}. The lattice that 
encodes the relationship $\preceq$ is known as the redundancy lattice (Fig.~\ref{fig:single_lat}), and guides the construction of the four terms in the PID.

\begin{figure}[ht]
  \centering
  \scalebox{0.9}{\includetikz{tikz/}{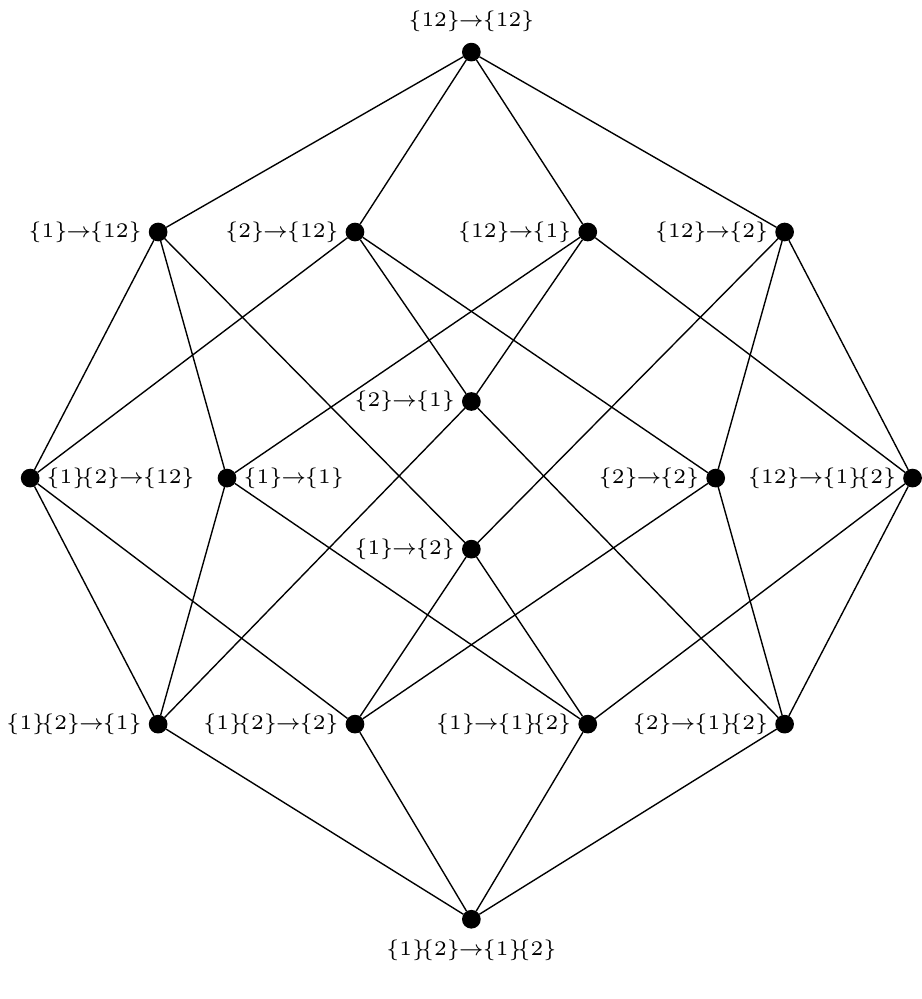}}
  \caption{The double-redundancy lattice for two predictors and two targets, 
  which is the product of two lattices as shown in Figure 2.}
  \label{fig:double_lat}
\end{figure}

Our first step is to build a \textit{product lattice} over $\mathcal{A}\times
\mathcal{A}$, in order to extend the notion of redundancy from PID to the case
of multiple source and target variables (here $X_1$, $X_2$ and $Y_1$, $Y_2$
respectively). Extending Williams and Beer's \cite{Williams2010} notation, we
denote sets of sources and targets using their indices only, with an arrow
going from past to future. Hence, the nodes of the product lattice are denoted
as $\bm \alpha \rightarrow \bm \beta$ for $\bm \alpha, \bm \beta \in
\mathcal{A}$, and a partial ordering relationship among them is defined by
\begin{equation}
\bm \alpha \rightarrow \bm \beta  \preceq \bm \alpha' \rightarrow \bm \beta' 
\quad \text{iff} \quad 
\bm \alpha \preceq \bm \alpha'
\:\: \text{and} \: \:
\bm \beta \preceq \bm \beta'.
\end{equation}
This relationship guarantees a lattice structure
\footnote{
A proof of this is provided in the Appendix. 
} 
with 16 nodes, which is shown in
Figure~\ref{fig:double_lat}. 
An intuitive understanding of the product lattice is developed in the sections below. 

\subsection*{Redundancies and atoms}

The next ingredient in the PID recipe is a \emph{redundancy function},
$I_\cap$, that quantifies the `overlapping' information about the target that
is common to a set of sources $\bm\alpha \in \mathcal{A}$~\cite{Williams2010}.
The redundancy function in PID includes the following terms:
$I_\cap^{\{1\}\{2\}}$ is the information about the target that is in either
source, $I_\cap^{\{i\}}$ the information in source $i$, and $I_\cap^{\{12\}}$
the information that is in both sources together. This subsection extends the
notion of overlapping information to the multi-target setting.

For a given $\bm \alpha \rightarrow \bm \beta \in \mathcal{A} \times
\mathcal{A}$, the overlapping information that is common to sources $\bm\alpha$
and can be seen in targets $\bm\beta$ is denoted as $\IR{\bm \alpha}{ \bm \beta
}$ and referred to as the \emph{double-redundancy function}. In the following,
we assume that the double-redundancy function satisfies two axioms:
\begin{itemize}
  \item \textbf{Axiom 1 (compatibility)}: if $\bm \alpha =\{\alpha_1,\dots,\alpha_J\}$
  and $\bm \beta =\{\beta_1,\dots,\beta_K\}$ with %
  $\bm\alpha,\bm\beta\in\mathcal{A}$ and $\alpha_j,\beta_k$ non-empty subsets of $\{1,\dots,N\}$, 
  then the following cases
  can be reduced to the redundancy of %
  PID or the mutual
  information~\footnote{We use the shorthand notation $\bm X^{\alpha} = (X_{i_1},\dots,X_{i_K})$ for
$\alpha = \{i_1,\dots,i_K\}$.}:%
\begin{equation}
\IR{\bm \alpha}{\bm \beta} = 
\begin{cases}
\texttt{Red}(\bm X^{\alpha_1},\dots, \bm X^{\alpha_J}; \bm Y^{\beta_1})
&\text{if}\quad K=1,\\
\texttt{Red}(\bm Y^{\beta_1},\dots, \bm Y^{\beta_K}; \bm X^{\alpha_1})
&\text{if}\quad J=1,\\
I(\bm X^{\alpha_1}; \bm Y^{\beta_1})
&\text{if}\quad J=K=1.
\end{cases}\nonumber
\end{equation}
  \item \textbf{Axiom 2 (partial ordering)}: if $\bm \alpha \rightarrow \bm
  \beta \preceq \bm \alpha' \rightarrow \bm \beta'$ then $\IR{\bm \alpha}{\bm
  \beta} \leq \IR{\bm \alpha'}{\bm \beta'}$.
\end{itemize}

By exploiting these axioms, one can define `atoms'
that belong to each of the nodes via the Moebius inversion formula. 
Concretely, the \textit{integrated information atoms} $\PI{\bm \alpha}{ \bm \beta}$ are defined
as the quantities that guarantee the following condition for all $
\bm \alpha \rightarrow \bm\beta \in \mathcal{A}\times \mathcal{A}$: 
\begin{align}
  \IR{\bm \alpha}{ \bm \beta} = \sum_{\substack{ \bm \alpha' \rightarrow\bm\beta' \preceq \bm\alpha\rightarrow\bm\beta}} \PI{\bm\alpha'}{\bm\beta'} ~.
  \label{eq:mi_decomp}
\end{align}
In other words, $\PI{\bm \alpha}{\bm \beta}$ corresponds to the information contained in node $\bm\alpha\rightarrow\bm\beta$ and not in any node below it in the lattice.
These are analogues to the redundant, unique, and synergistic atoms in the forward and backward PID above, but using the product lattice as a scaffold.
By inverting this relationship, one can find a recursive expression for calculating $I_{\partial}$ as
\begin{align}\label{eq:moebius}
  \PI{\bm \alpha}{\bm \beta} = \IR{\bm \alpha}{\bm \beta} - \sum_{\substack{ \bm \alpha' \rightarrow\bm\beta' \prec \bm\alpha\rightarrow\bm\beta}} \PI{\bm\alpha'}{\bm\beta'} ~ .
\end{align}
With all the tools at hand, we can deliver the promised decomposition of $\mathbf{E}$ in terms of atoms of integrated information.

\begin{definition}
The Integrated Information Decomposition ($\Phi$ID) of a system with Markovian dynamics is the collection of atoms $I_\partial$ defined from the redundancies $I_\cap$ via
Eq.~\eqref{eq:moebius}, which satisfy
\begin{align}
    \mathbf{E} = I(\bX; \bY) = \sum_{\substack{ \bm \alpha,\bm\beta \in \mathcal{A}}} \PI{\bm\alpha}{\bm\beta} ~ .
\end{align}
\end{definition}

It is direct to see that the $\Phi$ID of two time series gives 16 atoms that
correspond to the lattice shown in Figure~\ref{fig:double_lat}, which are
computed from a linear transformation over the 16 redundancies. Interestingly,
Axioms 1 and 2 allow us to compute all the $I_\cap$ terms once a single-target
PID redundancy function $\texttt{Red}(\cdot)$ has been chosen, with the sole
exception of $\IR{\{1\}\{2\}}{\{1\}\{2\}}$. All this is summarised in the
following result.

\begin{proposition}\label{prop:15forfree}
Axioms 1 and 2 provide unique values for the 16 atoms of the product lattice (see Figure~\ref{fig:double_lat}) after one defines (i) a single-target redundancy function $\texttt{Red}(\cdot)$, and (ii) an expression for $\PI{\{1\}\{2\}}{\{1\}\{2\}}$.
\end{proposition}
In the same way as in PID the definition of $\texttt{Red}(\cdot)$ 
gives 3 other terms (unique and synergy) as side-product, Proposition~\ref{prop:15forfree}
shows that in $\Phi \text{ID}$ the addition of the double-redundancy function 
$\PI{\{1\}\{2\}}{\{1\}\{2\}}$ gives 15 other terms for free~\footnote{Note that our framework does not prescribe a particular formula for $\PI{\{1\}\{2\}}{\{1\}\{2\}}$. A discussion on this issue can be found in the supplementary material.}. 

Throughout the rest of the article we outline how $\Phi$ID can be used to revise theories of information dynamics and integrated information, and how
it can provide more detailed analyses of systems of interest.

\subsection*{Simple examples}

To start developing our intuition about the $\Phi$ID atoms, let us decompose
the mutual information between the present of one variable, $X_i$, and its own
future, $Y_i$, i.e. the information storage in variable $i$ \cite{Lizier2010}:
\begin{align}
\begin{split}
I(X_i;Y_i) =&~ \PI{\{1\}\{2\}}{\{1\}\{2\}} + \PI{\{1\}\{2\}}{\{i\}} \\
&+\; \PI{\{i\}}{\{1\}\{2\}} +\PI{\{i\}}{\{i\}}~.%
\label{eq:ais}%
\end{split}
\end{align}
Here, \PI{\{1\}\{2\}}{\{1\}\{2\}} corresponds to redundant information in the
sources that is present in both targets; $\PI{\{1\}\{2\}}{\{i\}}$ is the
redundant information in the sources that is eliminated from the $j$-th source
($j\neq i$) and hence is only conserved in $Y_i$; and similarly for the
remaining atoms.

As another example, consider the transfer entropy from $i$ to $j$ (with $i \neq j$):
\begin{align}
\begin{split}
  I(X_i; Y_j|X_j) =&~ \PI{\{12\}}{\{1\}\{2\}} + \PI{\{12\}}{\{j\}} \\
  &+\; \PI{\{i\}}{\{1\}\{2\}} + \PI{\{i\}}{\{j\}}~.%
\label{eq:te}%
\end{split}
\end{align}
As before, $\PI{\{12\}}{\{1\}\{2\}}$ is the synergistic information present in
the joint past $(X_1,X_2)$ that can be read through either $Y_1$ or $Y_2$, and
similarly for the rest of the terms.

In the following section we explore the possibilities offered by this
decomposition, and its implications for causal analysis, IIT, and complex
systems in general.

\section*{Results}

\subsection*{Limitations of conventional causal discovery methods}

Mutual information and transfer entropy (or linear variants of them, to which
our conclusions also apply) are the building blocks of most popular methods of
statistical causal discovery. We now show that these metrics have two kinds of
limitations: they conflate multiple effects in counterintuitive ways, and they
fail to capture some effects altogether.

First, let us focus on the decomposition of information storage in
Eq.~\eqref{eq:ais}. Note that, although $X_2,Y_2$ are not in this mutual
information, $I(X_1;Y_1)$ shares the term $\PI{\{1\}\{2\}}{\{1\}\{2\}}$ with
$I(X_2;Y_2)$ by virtue of them being considered part of the same multivariate
stochastic process. Therefore, if one uses simple mutual information as a
measure of storage one may include information that is not stored exclusively
in a given variable, which may lead to paradoxical conclusions such as the sum
of individual storages being greater than $\mathbf{E}$.

Next, consider the terms in the decomposition of transfer entropy in
Eq.~\eqref{eq:te}. Note that, of these, $\PI{\{i\}}{\{j\}}$ is the only
`genuine' transfer term -- all others correspond to redundant or synergistic
effects involving both variables in past or future. Furthermore, one of the
`extra' terms ($\PI{\{i\}}{\{1\}\{2\}}$) is shared with $I(X_i;Y_i)$, in a
somewhat counterintuitive overlap between storage and transfer. Similar
concerns have been discussed in the literature \cite{james2016information},
showing that transfer entropy \emph{per se} cannot be taken as a pure measure
of information transfer.

Finally, from the decompositions of the mutual information and conditional
mutual information as shown above, it is clear that none of these quantities
are able to capture the $\Phi$ID terms of the form $\PI{\alpha}{\{12\}}$. These
terms correspond to `synergistic effects' (i.e. causes whose effects only
manifest on groups, rather than individual variables) and are neglected by
standard causal discovery methods.

\subsection*{Information processing in complex systems}

Based on $\Phi$ID, and building on Lizier's work~\cite{Lizier2010}, we propose
an extended taxonomy of information dynamics, with 6 disjoint and qualitatively
distinct phenomena:

\begin{description}

\item[Storage] Information that remains in the same source set, even if it
includes collective effects. Includes $\PI{ \{1\}\{2\} }{ \{1\}\{2\} }$, $\PI{
\{1\} }{ \{1\} }$, $\PI{ \{2\} }{ \{2\} }$, and $\PI{ \{12\} }{ \{12\} }$.

\item[Copy] Information that becomes duplicated. Includes $\PI{ \{1\} }{ \{1\}\{2\} }$, and
$\PI{ \{2\} }{ \{1\}\{2\} }$.

\item[Transfer] Information that moves between variables. Includes $\PI{ \{1\}
}{ \{2\} }$, and $\PI{ \{2\} }{ \{1\} }$.

\item[Erasure] Duplicated information that is pruned. Includes $\PI{
\{1\}\{2\} }{ \{1\} }$, and $\PI{ \{1\}\{2\} }{ \{2\} }$.

\item[Downward causation] Collective properties that define individual
futures. Includes $\PI{ \{12\} }{ \{1\} }$, $\PI{ \{12\} }{ \{2\} }$, and $\PI{
\{12\} }{ \{1\}\{2\} }$.

\item[Upward causation] Collective properties that are defined by individuals.
Includes $\PI{ \{1\} }{ \{12\} }$, $\PI{ \{2\} }{ \{12\} }$, and $\PI{
\{1\}\{2\} }{ \{12\} }$.

\end{description}

While downward causation has been discussed in the
past~\cite{james2016information}, upward causation and synergistic storage
($\PI{\{12\}}{\{12\}}$) have, to our knowledge, not been reported in the
literature. This revised taxonomy leads to less ambiguous, and more
quantifiable descriptions of information dynamics in complex systems, in
addition to grounding abstract concepts such as upward and downward
causation~\footnote{The relation between $\Phi$ID and causal
emergence~\cite{seth2010measuring} will be described in a separate
publication.}, and notions such as integrated information.

\subsection*{Different types of integration}

One important conceptual result of our framework is that there are multiple
qualitatively different ways in which a multivariate dynamical process can
integrate information through combinations of redundant, unique, or synergistic
effects. As elementary examples, consider the following systems of 2 binary
variables:

\begin{itemize}

  \item A \textbf{copy transfer} system, in which $x_1,x_2,y_1$ are
  i.i.d.  fair coin flips, and $y_2 = x_1$ (i.e.
  one bit is shifted).

  \item The \textbf{downward XOR}, in
  which $x_1,x_2,y_2$ are independent identically distributed fair coin flips, and $y_1 \equiv x_1 + x_2 ~\mathrm{(mod~2)}$.

  \item The \textbf{parity-preserving random} (PPR), in
  which $x_1,x_2$ are i.i.d. fair coin flips, and $x_1 + x_2
  \equiv y_1 + y_2~\mathrm{(mod~2)}$ (i.e. $\bm y$ is a random string of the same parity as $\bm x$).

\end{itemize}

\begin{figure}[ht]
  \centering
  \includetikz{tikz/}{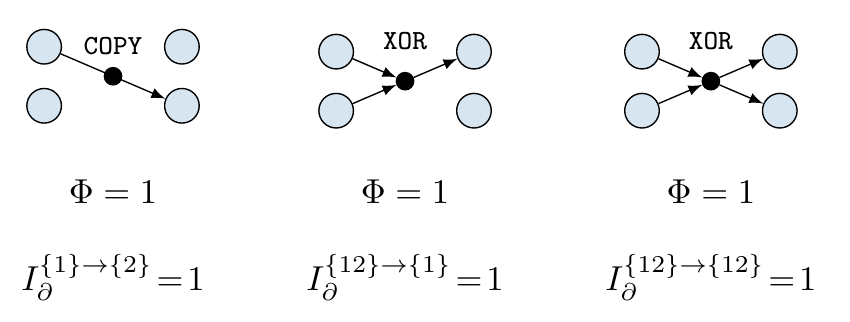}
  \caption{Example systems of logic gates. All of them have the same integrated
  information (measured with \wmsphi), but their information dynamics are
different. This difference is captured by a full $\Phi$ID decomposition,
that shows the only non-zero atoms are transfer (left), downward causation (centre),
and synergistic storage (right).}
  \label{fig:examples}
\end{figure}

These three systems (Fig.~\ref{fig:examples}) are `equally integrated,' in the
sense that the dynamics of the whole cannot be perfectly predicted from the
parts alone and the integrated information measure \wmsphi (later defined in
Eq.~\ref{eq:wmsphi}), yields $\wmsphi=1$ for all of
them~\cite{Mediano2019,Barrett2011}. However, they integrate information in
qualitatively different ways: in effect, the integration in the copy system is
entirely due to transfer dynamics (\PI{\{1\}}{\{2\}}); the downward
\texttt{XOR} integrates information due to downward causation
(\PI{\{12\}}{\{1\}}); and PPR due to synergistic storage (\PI{\{12\}}{\{12\}}).
All the other $\Phi$ID atoms in each of these systems are zero (proofs in the
Appendix).

\subsection*{Measures of integrated information}

Within the IIT literature, researchers have proposed multiple measures aimed at
quantifying to what extent the parts of a system affect each other's temporal
evolution. These measures, though superficially similar, are known to behave
inconsistently, for reasons that are not always clear~\cite{Mediano2019}. Here
we use $\Phi$ID to dissect and compare four existing measures of integrated
information ($\wmsphi$, $\psi$, $\Phi_G$) and dynamical complexity (CD). We do
not provide definitions of each measure here -- for details see Section 2.2 of
Ref.~\cite{Mediano2019} and the original
references~\cite{Balduzzi2008,Griffith2014,Oizumi2016}.

As a systematic exploration, one can determine which measures are sensitive to
which modes of information dynamics by calculating whether each measure is
zero, positive, or negative for a system consisting of only one particular
$\Phi$ID atom (Table~\ref{tab:comparison}; proofs in the Appendix). The main
result is that each measure captures a different combination of $\Phi$ID atoms:
although generally most of them capture synergistic effects and avoid (or
penalise) redundant effects, they differ substantially. The key conclusion is
that these measures are not simply different approximations of a unique concept
of integration, but that they are capturing intrinsically different aspects of
the system's information dynamics. While aggregate measures like these can be
empirically useful, one should keep in mind that they are measuring
combinations of different effects within the system's information dynamics.
Echoing the conclusions of Ref.~\cite{Mediano2019}: these measures behave
differently not only in practice, but also \emph{in principle}.

\vspace{-5pt}
\begin{table}[h]
  \caption{Sensitivity of integrated information measures to $\Phi$ID atoms. For each measure, entries indicate whether the value is positive (+), negative (-) or 0 in a system in which the given $\Phi$ID atom is the only non-zero atom.}
  \label{tab:comparison}
  \centering
  \renewcommand{\arraystretch}{1.2}
  \setlength\tabcolsep{5pt}
  \vspace{5pt}
  \begin{tabular}{l | c c c c}
    \multicolumn{1}{c}{\bfseries $\Phi$ID atoms} & \multicolumn{4}{c}{\textbf{Measures}} \\
    ~                                & $\Phi$ & \textrm{CD} & $\psi$ & $\Phi_G$ \\
    \toprule                                    
    \PI{\{1\}\{2\}}{\{1\}\{2\}}      &    -   &      0      &    0   &    0     \\
    \PI{\{1\}\{2\}}{\{i\}}           &    0   &      0      &    0   &    0     \\
    \PI{\{1\}\{2\}}{\{12\}}          &    +   &      0      &    0   &    0     \\
    \PI{\{i\}}{\{1\}\{2\}}           &    0   &      +      &    0   &    +     \\
    \PI{\{i\}}{\{i\}}                &    0   &      0      &    0   &    0     \\
    \PI{\{i\}}{\{j\}}                &    +   &      +      &    0   &    +     \\
    \PI{\{i\}}{\{12\}}               &    +   &      0      &    0   &    0     \\
    \PI{\{12\}}{\{1\}\{2\}}          &    +   &      +      &    +   &    +     \\
    \PI{\{12\}}{\{i\}}               &    +   &      +      &    +   &    +     \\
    \PI{\{12\}}{\{12\}}              &    +   &      0      &    +   &    0     \\
  \end{tabular}
\end{table}

\vspace{-9pt}

\subsection*{Why whole-minus-sum $\Phi$ can be negative}
\label{sec:wmsphi}

The $\Phi$ID can be further leveraged to provide elegant explanations of
certain behaviours of integrated information and dynamical complexity measures.
For example, $\Phi^{\mathrm{WMS}}$, which is calculated as
\begin{align}
  \Phi^{\mathrm{WMS}} = I(X_1,X_2; Y_1,Y_2) - I(X_1; Y_1) - I(X_2; Y_2)
  \label{eq:wmsphi}%
\end{align}
for a bivariate process, can sometimes take negative values. This feature,
which has been used as an argument to discard $\Phi^{\mathrm{WMS}}$ as a
suitable measure of integrated information \cite{Griffith2014,Oizumi2016}, can
be explained as follows. By applying the decomposition in
Eq.~\eqref{eq:mi_decomp}, one finds that
\begin{align*}
  \begin{rcases}\Phi^{\mathrm{WMS}} 
  =& - \PI{\{1\}\{2\}}{\{1\}\{2\}} \phantom{\texttt{Syn}(X_1,X_2; \!YY)} \end{rcases} \text{Red} \\
  \begin{rcases} 
  &+ \texttt{Syn}(X_1,X_2; Y_1Y_2) + \PI{\{1\}\{2\}}{\{12\}} \\
  &+ \PI{\{1\}}{\{12\}} + \PI{\{2\}}{\{12\}} \end{rcases} \text{Syn} \\
                 \begin{rcases} 
  &+ \PI{\{1\}}{\{2\}} + \PI{\{2\}}{\{1\}}~. \phantom{\texttt{Syn}(X;~)} \end{rcases} \text{Un\phantom{e}}
\end{align*}
Hence, $\Phi^{\mathrm{WMS}}$ accounts for all the synergies in the system
(the seven terms in Fig.~\ref{fig:double_lat} with $\{12\}$ in either side), the
unique information transferred between parts of the system, and, importantly,
the negative of the bottom node of the $\Phi$ID lattice. The presence of this negative
double-redundancy term shows that in highly redundant systems $\Phi^{\mathrm{WMS}}$ can be
negative. This is akin to Williams and Beer's \cite{Williams2010}
explanation of the negativity of the interaction information, applied to
multivariate processes. Based on this insight, one can formulate a `corrected'
$\Phi^{\mathrm{WMS}}$ by adding back the double-redundancy:
\begin{align*}
 \Phi^{\mathrm{WMS},c} :=  \Phi^{\mathrm{WMS}} + \PI{\{1\}\{2\}}{\{1\}\{2\}} ~ ,
\end{align*}
\noindent which includes only synergistic and unique transfer terms.

We computed $\Phi^{\mathrm{WMS},c}$ numerically for a simple example, using an
extension of the PID presented by James et al.~\cite{James2018}. Mimicking the
setting in Ref.~\cite{Mediano2019} with discrete variables, let us consider a
system in which $y_1,y_2$ are noisy AND gates of $x_1,x_2$ and the correlation
between the noise components of $y_1$ and $y_2$ is a free parameter. We
calculated $\Phi^{\mathrm{WMS}}$ and $\Phi^{\mathrm{WMS},c}$ with respect to
the system's stationary distribution. Plots of the standard and corrected
$\Phi^{\mathrm{WMS}}$ for this system are shown in Fig,~\ref{fig:wms_ar}, and
details of the computation can be found in the Appendix.

\begin{figure}[ht]
  \centering
  \includetikz{tikz/}{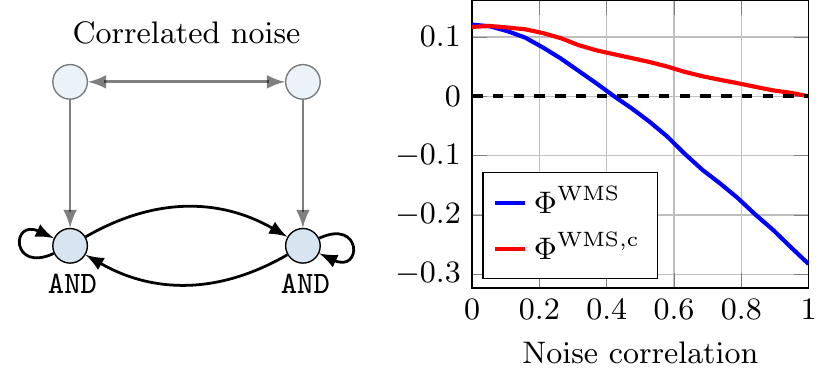}

  \caption{Standard and corrected \wmsphi in a two component noisy AND system with varying correlation in the noise to each component.}

  \label{fig:wms_ar}
\end{figure}

As expected, \wmsphi drops below zero as synergy decreases and redundancy
increases with noise correlation. Interestingly, after adding the
double-redundancy term, the corrected version, $\Phi^{\mathrm{WMS},c}$, tends
to 0 for high noise correlation, which is more similar to some of the other
measures highlighted in \cite{Mediano2019}, e.g.~CD and $\Phi^*$.

\subsection*{Why unnormalised causal density can exceed TDMI}
\label{sec:ucd}

\footnotetext[999]{Note that
the original definition of causal density is normalised by $L(L-1)$, and has
been proven to be bounded by mutual information \cite{Mediano2019}.}

In Oizumi et al. \cite{Oizumi2016}, the authors correctly point out that the
sum of conditional pairwise transfer entropies (or unnormalised Causal Density;
uCD) in a system can exceed the total mutual information, which is problematic
for considering this as a measure of integrated
information~\cite{Oizumi2016,Note999}. This quantity, given by
\begin{align}
  \mathrm{uCD} &= \mathrm{TE}_{Z_1 \rightarrow Z_2} + \mathrm{TE}_{Z_2 \rightarrow Z_1} \nonumber \\
              &= I(X_1; Y_2|X_2) + I(X_2; Y_1|X_1) ~ ,
  \label{eq:ucd}
\end{align}
can also be decomposed using $\Phi$ID. By applying Eq.~\eqref{eq:mi_decomp} to
the expression of uCD, one finds that
\begin{align*}
\mathrm{uCD} =& \texttt{Un}(X_1; Y_2 | X_2) + \PI{\{12\}}{\{2\}} \\
+& \texttt{Un}(X_2; Y_1|X_1) + \PI{\{12\}}{\{1\}} \\
+& 2 \PI{\{12\}}{\{1\}\{2\}} ~ .
\end{align*}
Besides the unique and synergistic terms that one would expect in a measure of
information transfer \cite{Williams2011}, there is in addition a
double-counting of a downward causation $\Phi$ID atom,
$\PI{\{12\}}{\{1\}\{2\}}$. Specifically, uCD double-counts synergistic
information in the past that is transferred redundantly to the future, and this
can cause uCD to be greater than $I(X_1, X_2; Y_1,Y_2)$.

This finding makes it straightforward to design systems for which uCD is maximal 
(i.e. a system that has only $\PI{\{12\}}{\{1\}\{2\}}>0$): $x_1,x_2$ are maximum 
entropy and $y_1 = y_2 = x_1 \oplus x_2$. Indeed, for this system 
$\mathrm{uCD} = 2\text{~bit} > I(X_1,X_2; Y_1,Y_2) = 1\text{~bit}$.

Furthermore, this decomposition also shows that there are many common atoms in
the $\Phi$ID expansions of CD and $\Phi^{\text{WMS}}$, which might explain why
CD has sometimes been considered together with measures of integrated
information~\cite{Seth2011,Mediano2019}.

\section*{Discussion}

We propose $\Phi$ID as a novel information-theoretic framework to study
high-order interactions in time-series data. By unifying aspects of integrated
information theory (IIT) and partial information decomposition (PID), the
$\Phi$ID framework allows us to decompose information flow in a multivariate
stochastic process into interpretable, disjoint parts. This allows systematic
studies of unexplored modes of information dynamics -- including modes of
synergistic storage, and upward and downward causation -- in a purely
data-driven fashion.

\subsection*{Towards multi-dimensional measures of complexity}

Besides the importance of having an encompassing taxonomy of information
dynamics phenomena, this frameworks suggests, following Feldman and Crutchfield
\cite{Feldman1998}, that there is no theoretical basis to a purported
all-encompassing scalar measure of dynamical complexity. The richness of
complex dynamics is vast, and the prospect of subsuming all into a single
number is unreasonable. Scalar measures might still have great practical value
in certain contexts~\footnote{For example, measures that accurately
discriminate between neural configurations corresponding to conscious and
unconscious states in a particular experimental paradigm~\cite{Casali2013}.};
nevertheless, a general theory of complex systems (biological or otherwise)
cannot be reduced to a single, one-size-fits-all measure.

\subsection*{Integration measures conflate transfer and synergy}

Using $\Phi$ID, one is able to inspect previous measures of integrated
information, explaining similarities and differences between them, and fixing
some of their shortcomings. Most importantly, we have shown that what is
usually referred to as `integration' is in fact an aggregate of several
different information effects, typically including transfer and synergy
phenomena. Moreover, different measures capture different effects in various
proportions, which explains the heterogeneity among existing measures reported
in Ref.~\cite{Mediano2019}. By employing $\Phi$ID one can tailor measures for
targeting specific mixtures of effects, according to the information dynamics
processes one wishes to analyse.

\subsection*{Causal analysis}

\footnotetext[998]{Intuitively, a causal analysis reveals what the system \emph{could do}, while a dynamical analysis based on attractor statistics reveals what the system \emph{actually does}.}

As presented, $\Phi$ID is a generic tool to decompose multivariate mutual
information, which can be directly used to perform causal analysis. Most
integrated information measures can be roughly divided between those that
describe integration in a system based on its causal
properties~\cite{Oizumi2014}, and those that use the system's attractor
statistics, known as \emph{dynamical} integration
measures~\cite{Mediano2019,Note998}. Given a system's conditional probability
distribution $p(\bm Y | \bm X)$, one can use $\Phi$ID to perform either a
causal or a dynamical analysis by using the stationary attractor distribution
$p(\bX)$, or a maximum entropy distribution on $\bm X$. However, note that a
few additional assumptions need to hold to interpret the results in a strict
causal sense; in particular, the conditional distribution $p(\bY|\bX)$ needs to
be equivalent to a \texttt{do()} distribution in Pearl's
sense~\cite{pearl2018book}, and the system must satisfy the faithfulness and
causal Markov conditions~\footnote{Also, note that the maximum entropy
distributions employed by some causal integration frameworks are well-defined
for discrete Markovian systems, but in general may not always exist
\cite{Barrett2019}.}.

\subsection*{Limitations and future extensions}

Our method inherits some of the limitations of PID. In particular, several
distinct redundancy functions have been proposed for evaluating PID atoms, but
there is not yet a consensus on one that is universally
preferable~\cite{James2018}. Forthcoming work will explore how the \phiid
framework yields new dynamical insights into redundancy function selection, and
helps us address the current challenges of PID.

\begin{acknowledgments}

The authors thank Julian Sutherland for valuable discussions. F.E.R. was
supported by the Ad Astra Chandaria Foundation, and by the European Union’s
H2020 research and innovation programme under the Marie Sk\l{}odowska-Curie
grant agreement No. 702981. A.K.S. and A.B.B. acknowledge support from the Dr
Mortimer and Theresa Sackler Foundation. A.K.S. also acknowledges support from
the Canadian Institute for Advanced Research (CIFAR): Azrieli Programme on
Brain, Mind, and Consciousness.

\end{acknowledgments}

\appendix

\section{The product of two lattices is a lattice}

A lattice is a partially ordered set $(\mathcal{A}, \preceq)$ for which every
pair of elements $a,b$ has a well-defined \textit{meet} $a \wedge b$ and
\textit{join} $a \vee b$, which correspond to their common greatest lower bound
(infimum) and common least upper bound (supremum),
respectively~\cite{charalambides2002enumerative}. Here we prove that, if
$(\mathcal{A},\preceq)$ is a lattice, then the product lattice $(\mathcal{A}
\times \mathcal{A}, \preceq^*)$ equipped with the order relationship
\begin{equation}
\alpha \rightarrow \beta \preceq^* \alpha' \rightarrow \beta' 
\quad \text{if and only if} \quad 
\alpha \preceq \alpha'
\:\: \text{and} \: \:
\beta \preceq \beta',
\end{equation}
\noindent is also a lattice, where $\alpha, \beta, \alpha', \beta' \in
\mathcal{A}$. As a corollary of this, given that the set and partial ordering
relationship used in PID are a lattice \cite{Williams2010,Crampton2001}, then
the set and partial ordering relationship used in $\Phi$ID are also a lattice.

For compactness, let us use the notation $\gamma = \alpha \rightarrow \beta$
and $\gamma' = \alpha' \rightarrow \beta'$ for $\gamma,\gamma'\in
\mathcal{A}\times\mathcal{A}$. To prove the lattice structure of $(\mathcal{A}
\times \mathcal{A}, \preceq^*)$ it suffices to show that

\begin{enumerate}
    \item $\gamma \meet^* \gamma' := \alpha \meet \alpha' \rightarrow \beta \meet \beta'$ is a valid meet; and
    \item $\gamma \join^* \gamma' := \alpha \join \alpha' \rightarrow \beta \join \beta'$ is a valid join.
\end{enumerate}
Note that the fact that $(\mathcal{A},\preceq)$ is a lattice implies that
$\alpha \meet \beta$ and $\alpha \join \beta$ are well-defined for all
$\alpha,\beta\in\mathcal{A}$.

Let us begin with the meet, for which we use $m = \gamma \meet^* \gamma'$ as a
shorthand notation. First, one can directly check that $m \preceq^* \gamma$ and
$m \preceq^* \gamma'$, given the definition of $\preceq^*$ above and the fact
that $\alpha \meet \alpha' \preceq \alpha$ (and similarly for $\alpha'$,
$\beta$, and $\beta'$). Next, we need to prove that for any $\gamma'' =
\alpha'' \rightarrow \beta'' \in \mathcal{A} \times \mathcal{A}$ such that
$\gamma'' \preceq^* \gamma$ and $\gamma'' \preceq^* \gamma'$, we have $\gamma''
\preceq^* m$ (i.e. that $m$ is the greatest lower bound of $\gamma$ and
$\gamma'$). To see this, note that the conditions $\gamma'' \preceq^* \gamma$
and $\gamma'' \preceq^* \gamma'$ imply the following four statements:
\begin{align*}
    \alpha'' & \preceq ~ \alpha ~, \\
    \alpha'' & \preceq ~ \alpha' ~,\\
    \beta''  & \preceq ~ \beta ~,\\
    \beta''  & \preceq ~ \beta' ~ .
\end{align*}
Using these relationships and the $\meet$ operator from $\mathcal{A}$, one can
show that $\alpha'' \preceq \alpha \meet \alpha'$ and $\beta'' \preceq \beta
\meet \beta'$, which in turn implies that $\gamma'' \preceq^* m$. Finally, the
proof for the join is analogous, replacing $\meet$ with $\join$ and $\preceq$
with $\succeq$.

\section{Decomposing PID atoms}

Equation (4) in the main text shows how to decompose redundancies in the
product lattice in terms of $\Phi$ID atoms. Here we provide a more general
statement, that allows us to decompose not only redundancies, but also other
PID atoms. The goal of this appendix is to build stronger connections between
PID and \phiid, and to extend Proposition 1 to allow greater flexibility for
specifying a \phiid function.

For the forward PID, and borrowing the notation from Williams and Beer
\cite{Williams2010}, given a non-empty set of `future' variables $F \in
\mathcal{P}(\{Y_1, ... , Y_N\})$ and an an element of the redundancy lattice
$\alpha \in \mathcal{A}$, let us denote by $\Pi_F(\alpha; F)$ the $\alpha$ atom
of the PID decomposition for $I(\bX; F)$, such that
\begin{align}
    I(\bX; F) = \sum_{\alpha \in \mathcal{A}} \Pi_F(\alpha; F) ~ .
\end{align}
We use an analogous notation for the backward PID, with a corresponding
non-empty set of `past' variables $P \in \mathcal{P}(\{X_1, ... , X_N\})$ and
$\beta \in \mathcal{A}$, such that
\begin{align}
    I(P; \bY) = \sum_{\beta \in \mathcal{A}} \Pi_B(P; \beta) ~ .
\end{align}

Then, these quantities can be further decomposed in \phiid atoms as
\begin{subequations}
\begin{gather}
    \Pi_F(\alpha; F) = \sum_{\gamma \preceq F} \PI{\alpha}{\gamma} ~ , \label{eq:pid_decomp} \\
    \Pi_B(P;  \beta) = \sum_{\gamma \preceq P} \PI{\gamma}{\beta} ~ .
\end{gather}
\end{subequations}

Note that the sum runs only across one of the sets (instead of both as it does
in Eq.~(4) of the main text), and that every element in $\mathcal{P}(\{1, ... ,
N\})$ is also in $\mathcal{A}$, and hence the partial order relationship in the
sums above is well-defined. As a few examples, in a bivariate system the
following forward PID atoms decompose as:
\begin{align*}
    \texttt{Red}(X_1, X_2; Y_i) &= \Pi_F(\{1\}\{2\}; Y_i) \\
    &= \PI{\{1\}\{2\}}{\{1\}\{2\}} + \PI{\{1\}\{2\}}{\{i\}}~, \\[2ex]
    \texttt{Syn}(X_1, X_2; Y_i) &= \Pi_F(\{12\}; Y_i) \\
                                &= \PI{\{12\}}{\{1\}\{2\}} + \PI{\{12\}}{\{i\}} ~,\\[2ex]
    \texttt{Un}(X_1; Y_1Y_2 | X_2)  &= \Pi_F(\{1\}; Y_1Y_2) \\
                                &= \PI{\{1\}}{\{1\}\{2\}} + \PI{\{1\}}{\{1\}} \\
                                &~ ~ ~+ \PI{\{1\}}{\{2\}} + \PI{\{1\}}{\{12\}}~.
\end{align*}

These decompositions can be used to prove Proposition~1 of the main text.
Adopting a view of $\Phi$ID as a linear system of equations, one needs 16
independent equations to solve for the 16 unknowns that are the \phiid atoms.
Of those, 9 are given by standard Shannon mutual information (specifically,
$I(X_i;Y_j)$, $I(X_1X_2; Y_i)$, $I(Y_1Y_2; X_i)$, and $I(X_1X_2; Y_1Y_2)$, for
$i,j=\{1,2\}$) decomposed with Eq.~(4) of the main text, and 6 are given by the
single-target PIDs ($\texttt{Red}(X_1, X_2; Y_1)$, $\texttt{Red}(X_1, X_2;
Y_2)$, and $\texttt{Red}(X_1, X_2; Y_1Y_2)$, as well as the 3 corresponding
backward PIDs) decomposed by the expression above. Finally, one only need to
add one individual \phiid atom to make the 16 equations needed, and the system
can be solved for all other atoms.

Taking these results together, Proposition~1 in the main text can be
generalised as follows: a valid $\Phi$ID can be defined not only in terms of
redundancy, but also in terms of unique information or synergy. This is
equivalent to the case of PID, for which decompositions based on unique
information \cite{James2018} or synergy
~\cite{quax2017quantifying,rassouli2018latent} have been proposed. In fact, for
the numerical results in Fig.~5 of the main text we use a $\Phi$ID based on
unique information defined below.

\section{Computing the $\Phi$ID atoms}

In Ref.~\cite{James2018}, James, Emenheiser and Crutchfield introduce a PID
based on a new measure of unique information, $I_{\mathrm{dep}}$, which we
succinctly describe here. To define $I_{\mathrm{dep}}$, they first define a
\emph{constraint lattice} $\mathcal{L}$ on a set of variables (formally defined
as the set of antichain covers with the natural partial ordering).
Specifically, given a constraint $\sigma$ and a probability distribution $p$,
consider the set $\Delta_p(\sigma)$ of distributions that match marginals in
$\sigma$ with $p$:
\begin{align*}
    \Delta_p(\sigma) = \{q: p(\gamma) = q(\gamma), \gamma \in \sigma \}~.
\end{align*}
For example, the constraint $\sigma = \{(X,Y), (X,Z)\}$ determines the set of
distributions $q$ such that $q(x,y) = p(x, y)$ and $q(x,z) = p(x, z)$. In
addition, the elements of $\mathcal{L}$ (i.e. the nodes in the lattice) have an
associated value of an information-theoretic measure $f[p_\sigma]$ evaluated on
$p_\sigma= \arg \max\{H[q] : q \in \Delta_p(\sigma)\}$.

Let us focus on the bivariate PID: denote by $L$ the collection of edges of the
constraint lattice for the variables $X,Y,Z$, and let $f$ be the joint mutual
information $I(XY; Z)$. For a link $(\sigma_1,\sigma_2)\in L$, one can evaluate
the change in $f$ along the link via the operator
$\Delta^{\sigma_1}_{\sigma_2}$; e.g. $\Delta^{\sigma_1}_{\sigma_2} I(XY;Z) =
I_{\sigma_1}(XY;Z) - I_{\sigma_2}(XY;Z) $. Additionally, for any $\gamma\in
\mathcal{P}(\{X,Y,Z\})$ let us define $E(\gamma)$ to be the set of all links
that contain $\gamma$ only at one side, i.e.
\begin{equation}
E(\gamma) = \{ (\sigma_1,\sigma_2) \in L : \gamma \in \sigma_1, \gamma \notin \sigma_2 \}.
\end{equation}
Then, the unique information is defined by
\begin{equation}
I_\text{dep}(X\to Z|Y) = \min_{(\sigma_1,\sigma_2)\in E(X,Z)} \Delta^{\sigma_1}_{\sigma_2} I(XY;Z) ~ .
\label{eq:pid_unique}%
\end{equation}
That is, the unique information is the smallest perturbation that is seen when
adding the dependency between $X$ and $Z$. For further details, and a more
pedagogical introduction, we refer the reader to the original paper
\cite{James2018}.

This measure can be naturally generalized to the $\Phi$ID setting by replacing
$I(XY; Z)$ above with the full joint mutual information $I(\bX; \bY)$ and
formulating the appropriate constraint lattice for $(\bX, \bY)$. More
precisely:
\begin{definition}
\textbf{Double-unique information based on dependencies}. For a given set of
variables $(\bm X,\bm Y)$, and two indices $i$ and $j$, the double-unique
information based on dependencies is defined as
\begin{equation}
I^{\{i\}\to\{j\}}_{\partial,\mathrm{dep}} := \min_{(\sigma_1,\sigma_2)\in E(X_i,Y_j)} \Delta^{\sigma_1}_{\sigma_2} I(\bm X; \bm Y).
\label{eq:phiid_double_unique}
\end{equation}
\end{definition}

This definition is applicable to both discrete and continuous random variables.
In practice, the difficulty of calculating $I_{\mathrm{dep}}$ amounts to the
difficulty of calculating maximum-entropy projections, which for Gaussian and
discrete distributions is easily done with off-the-shelf software -- in the
case of discrete variables, for example using the \texttt{dit} package
\cite{james2016information}. Once the double-unique information has been
calculated, the same lattice can be reused to compute the unique information
atoms for all 6 single-target PIDs, and together with the 9 MIs, these 16
numbers fully determine the numerical values of every $\Phi$ID atom.

It is important to recall that, as mentioned in the main body of the paper, the
two axioms of $\Phi$ID do not uniquely determine
$I^{\{i\}\to\{j\}}_{\partial}$. An exploration of alternative decompositions
will be covered in a separate publication.

\section{Results of section `Different types of integration'}

Here we present calculations for the example systems in Fig. 4 of the main
text. These proofs hold for all $\Phi$ID that satisfy the partial ordering
axiom of $\IR{\alpha}{\beta}$ (Axiom 2 in the main text), have a non-negative
double-redundancy function $I^{\{1\}\{2\}\to\{1\}\{2\}}\geq 0$, and satisfy the
following bound that follows from the basic properties of PID
(c.f.~\cite{rosas2016understanding}):
\begin{align}
  \texttt{Red}(X,Y;Z) \leq \min\{ I(X;Z), I(Y;Z) \} ~ .
\end{align}
Let us examine the three systems in turn:

\begin{itemize}

\item For the copy transfer system, $Y_2=X_1$, while $X_2$ and $Y_2$ are
independent i.i.d. fair coin flips. Since $Y_2$ is independent from the rest of
the system, $\texttt{Red}(X_1,X_2;Y_2) = \texttt{Red}(X_1,X_2;Y_2) = 0$, and
due to partial ordering $\IR{\{1\}\{2\}}{\{1\}\{2\}} = 0$. Finally, using the
Moebius inversion formula it follows that $\PI{\{1\}}{\{2\}} = I(X_1;Y_2) = 1$
and all other atoms are zero.

\item In the downward XOR system, $X_1$ and $X_2$ are i.i.d. fair coin flips,
$Y_1 = X_1 \oplus X_2$, and $Y_2$ is independent of the rest. Then, it is clear
that $I(X_1,X_2;Y_1,Y_2) = I(X_1,X_2;Y_1) = 1$, while
$I(X_1;Y_1)=I(X_2;Y_1)=0$. Additionally, note that $\IR{\{12\}}{\{1\}\{2\}}=0$,
since $\texttt{Red}(Y_1,Y_2;X_1 X_2) \leq I(Y_2; X_1 X_2) = 0$. All this
implies that all the redundancies (and hence all the atoms) below
$\{12\}\to\{1\}$ are zero, and hence $\PI{\{12\}}{\{1\}}=1$ due to the Moebius
inversion formula.

\item Finally, consider the PPR system where $X_1,X_2,Y_1$ are i.i.d. fair coin
flips and $Y_2$ is such that $X_1\oplus X_2 = Y_1\oplus Y_2$. Then
$I(X_1,X_2;Y_1) = I(X_1,X_2;Y_2) = I(X_1;Y_1,Y_2)=I(X_2;Y_1,Y_2)=0$. This
implies that all redundancies (and hence atoms) except $\IR{\{12\}}{\{12\}}$
are zero, and hence using again the Moebius inversion formula
$\PI{\{12\}}{\{12\}}=I(X_1,X_2;Y_1,Y_2)=1$.

\end{itemize}

\section{Results of section `Measures of integrated information'}

In this appendix we prove the results in Table 1 of the main text, that shows
whether each of four measures of integrated information ($\wmsphi$, CD, $\psi$,
$\Phi_G$) are positive, negative, or zero in a system containing only one
\phiid atom. A succinct definition of each measure is given below, and a
comprehensive review and comparison of these and other measures can be found in
Ref.~\cite{Mediano2019}.

Throughout this section we focus on bivariate systems, and use $i,j$ as
variable indices, with $i \neq j$. To complete the proof we will first show
that it is possible to build systems with exactly one bit of information in one
\phiid atom, and we will then compute the four measures on those systems.

Let us begin with the design of systems with one specific \phiid atom.
Intuitively, this can be accomplished with a suitable combination of COPY and
XOR gates for redundant and synergistic sets of variables, respectively. More
formally, the procedure to build a system with $\PI{\alpha}{\beta} = 1$ and all
other atoms equal to zero is as follows:

\begin{enumerate}
    \item Sample $w$ from a Bernoulli distribution with $p = 0.5$.
    \item Sample $\bm x$ based on $\alpha$:
\begin{itemize}
    \item If $\alpha = \{1\}\{2\}$, then $x_1 = x_2 = w$.
    \item If $\alpha = \{i\}$, then $x_i = w$ and $x_j$ is sampled from a Bernoulli distribution with $p = 0.5$.
    \item If $\alpha = \{12\}$, then $\bm x$ is a random string with parity $w$.
\end{itemize}
    \item Sample $\bm y$ based on $\beta$ analogously.
\end{enumerate}

In all cases there will be one bit of information ($w$) shared between $\bX$
and $\bY$, hence $I(\bX; \bY) = 1$ for any choice of $\alpha, \beta$. This can
be proven using the fact that for any $\alpha, \beta$, one has $H(W) = 1$, $H(W
| \bm X) = H(W | \bm Y) = 0$, and $p(\bm x, \bm y, w) = p(\bm x|w)p(\bm
y|w)p(w)$. To do so, let us start from the mutual information chain rule:
\begin{align*}
    I(\bm X; \bm YW) &= I(\bm X; W) + I(\bm X; \bm Y|W) \\
             &= I(\bm X; \bm Y) + I(\bm X; W|\bm Y) ~ .
\end{align*}
Rearranging the above terms, one can find that
\begin{align*}
    I(\bm X; \bm Y) = I(\bm X; W) + I(\bm X; \bm Y|W) - I(\bm X; W|\bm Y) ~ ,
\end{align*}
where $I(\bm X; W) = H(W) - H(W|\bm X) = 1$ and $I(\bm X; \bm Y|W) = 0$.
Finally, one finds that
\begin{align*}
    I(\bm X; W|\bm Y) &= H(\bm X|\bm Y) + H(W|\bm Y) - H(\bm XW|\bm Y) \\
    &= H(\bm X|\bm Y) + H(W|\bm Y) ~ - \\ & \qquad \left( H(\bm X|\bm Y) + H(W|\bm X\bm Y) \right) = 0 ~ ,
\end{align*}
which concludes the proof that $I(\bm X; \bm Y) = 1$. Furthermore, following a
procedure similar to those in the previous section, it can be shown that any
\phiid that satisfies the axioms described above (partial ordering,
non-negative double-redundancy, and upper-bounded redundancy) correctly assigns
1 bit of information to $\PI{\alpha}{\beta}$, and 0 to all other atoms.

Now that we have built these 16 single-atom systems, let us move to the
integration measures of interest. For CD, $\psi$, and \wmsphi, we will proceed
by decomposing them in terms of \phiid atoms and checking whether each atom is
positive (+), negative (--), or absent (0) from the decomposition to obtain the
results in Table 1 of the article. Let us begin with CD, defined as the sum of
transfer entropies from one variable to the other:
\begin{align}
\begin{split}
    \mathrm{CD} &= \frac{1}{2} \sum_{i=1}^2 I(X_i; Y_j | X_j) \\
                &= \frac{1}{2} \sum_{i=1}^2 \Big( \PI{\{i\}}{\{1\}\{2\}} + \PI{\{i\}}{\{j\}} \\ & \qquad \qquad + \PI{\{12\}}{\{1\}\{2\}} + \PI{\{12\}}{\{j\}} \Big) ~ .
\end{split}
\end{align}
Similarly, for $\psi$ the atoms can be extracted from the decomposition of
$\texttt{Syn}(X_1, X_2; Y_1 Y_2)$ in Eq.~\eqref{eq:pid_decomp}:
\begin{align}
\begin{split}
    \psi &= \texttt{Syn}(X_1, X_2; Y_1 Y_2) \\ &= \PI{\{12\}}{\{1\}\{2\}} + \PI{\{12\}}{\{1\}} + \PI{\{12\}}{\{2\}} + \PI{\{12\}}{\{12\}} ~ .
\end{split}
\end{align}
For \wmsphi, the atoms can be extracted from the decomposition of Eq.~(9) in
the main text:
\begin{align}
\begin{split}
    \wmsphi = & -\PI{\{1\}\{2\}}{\{1\}\{2\}} + \PI{\{1\}\{2\}}{\{12\}} \\
      &+ \psi + \sum_{i=1}^2 \left( \PI{\{i\}}{\{j\}} + \PI{\{i\}}{\{12\}} \right) ~ .
\end{split}
\end{align}

The $\Phi_G$ case is slightly more involved, since it is not easily
decomposable into a sum of \phiid atoms. According to the definition of
$\Phi_G$ \cite{Oizumi2016}, for a system given by the joint probability
distribution $p(\bX, \bY)$ one has
\begin{align*}
    \Phi_G = \min_{q \in \mathcal{M}_G} D_{\mathrm{KL}}(p \| q) ~ ,
\end{align*}
where $\mathcal{M}_G$ is the manifold of probability distributions that satisfy
the constraints
\begin{align}
    q(Y_i | \bX) = q(Y_i | X_i) ~ .
    \label{eq:phig_constraint}%
\end{align}
Therefore, it suffices to check whether the probability distribution of the
system satisfies the constraints in Eq.~\eqref{eq:phig_constraint} --- if it
does, then $\Phi_G = 0$, and otherwise $\Phi_G > 0$ ---, which can be easily
verified for each system separately to obtain the $\Phi_G$ column in Table 1,
concluding the proof.

\vspace{20pt}

\section{Results of section `Why whole-minus-sum $\Phi$ can be negative'}

In this appendix we describe the details of the noisy AND system and how to
compute its \phiid to yield the results shown in Figure 4 of the main text.

Given the past state of the system $x_1x_2$, the next state is given by
\begin{align*}
    y_1 &= (x_1 \cdot x_2) \oplus n_1 \\
    y_2 &= (x_1 \cdot x_2) \oplus n_2 ~ ,
\end{align*}
where $n_1,n_2$ are two auxiliary noise variables sampled from Bernoulli
distributions with parameter $p = 0.2$, and they are sampled independently with
probability $1-c$ and set to be identical to each other with probability $c$.
This results in a system that, for $c = 0$, consists of two separate AND gates
with some noise, and for $c = 1$ a system of two perfectly correlated
components that at each time step change state with probability $0.2$. All
information-theoretic functionals are computed with respect to the system's
stationary distribution.

To compute the \phiid atoms we follow the procedure described above based on
James et al.'s $I_{\mathrm{dep}}$ measure. To minimise numerical problems with
the maximum-entropy projections involved, instead of computing all relevant
quantities separately we compute one single constraint lattice for the whole
system $X_1X_2Y_1Y_2$ and read off all relevant quantities:
\begin{itemize}
    \item 9 values of mutual information, which can be directly read from the corresponding nodes in the lattice;
    \item 6 values of single-target PID unique information, which can be obtained as the minimum of suitable subsets of the lattice according to Eq.~\eqref{eq:pid_unique}; and
    \item One \phiid double-unique information according to Eq.~\eqref{eq:phiid_double_unique}.
\end{itemize}
Together, these 16 numbers fully determine all 16 \phiid atoms, and the resulting linear system of equations can be easily solved.

\bibliography{main.bib}

\begin{thebibliography}{43}%
\makeatletter
\providecommand \@ifxundefined [1]{%
 \@ifx{#1\undefined}
}%
\providecommand \@ifnum [1]{%
 \ifnum #1\expandafter \@firstoftwo
 \else \expandafter \@secondoftwo
 \fi
}%
\providecommand \@ifx [1]{%
 \ifx #1\expandafter \@firstoftwo
 \else \expandafter \@secondoftwo
 \fi
}%
\providecommand \natexlab [1]{#1}%
\providecommand \enquote  [1]{``#1''}%
\providecommand \bibnamefont  [1]{#1}%
\providecommand \bibfnamefont [1]{#1}%
\providecommand \citenamefont [1]{#1}%
\providecommand \href@noop [0]{\@secondoftwo}%
\providecommand \href [0]{\begingroup \@sanitize@url \@href}%
\providecommand \@href[1]{\@@startlink{#1}\@@href}%
\providecommand \@@href[1]{\endgroup#1\@@endlink}%
\providecommand \@sanitize@url [0]{\catcode `\\12\catcode `\$12\catcode
  `\&12\catcode `\#12\catcode `\^12\catcode `\_12\catcode `\%12\relax}%
\providecommand \@@startlink[1]{}%
\providecommand \@@endlink[0]{}%
\providecommand \url  [0]{\begingroup\@sanitize@url \@url }%
\providecommand \@url [1]{\endgroup\@href {#1}{\urlprefix }}%
\providecommand \urlprefix  [0]{URL }%
\providecommand \Eprint [0]{\href }%
\providecommand \doibase [0]{http://dx.doi.org/}%
\providecommand \selectlanguage [0]{\@gobble}%
\providecommand \bibinfo  [0]{\@secondoftwo}%
\providecommand \bibfield  [0]{\@secondoftwo}%
\providecommand \translation [1]{[#1]}%
\providecommand \BibitemOpen [0]{}%
\providecommand \bibitemStop [0]{}%
\providecommand \bibitemNoStop [0]{.\EOS\space}%
\providecommand \EOS [0]{\spacefactor3000\relax}%
\providecommand \BibitemShut  [1]{\csname bibitem#1\endcsname}%
\let\auto@bib@innerbib\@empty
\bibitem [{\citenamefont {Kelso}(1995)}]{kelso1995dynamic}%
  \BibitemOpen
  \bibfield  {author} {\bibinfo {author} {\bibfnamefont {J.~S.}\ \bibnamefont
  {Kelso}},\ }\href@noop {} {\emph {\bibinfo {title} {Dynamic patterns: The
  self-organization of brain and behavior}}}\ (\bibinfo  {publisher} {MIT
  press},\ \bibinfo {year} {1995})\BibitemShut {NoStop}%
\bibitem [{\citenamefont {Runge}\ \emph {et~al.}(2019)\citenamefont {Runge},
  \citenamefont {Bathiany}, \citenamefont {Bollt}, \citenamefont {Camps-Valls},
  \citenamefont {Coumou}, \citenamefont {Deyle}, \citenamefont {Glymour},
  \citenamefont {Kretschmer}, \citenamefont {Mahecha}, \citenamefont
  {Mu{\~n}oz-Mar{\'\i}} \emph {et~al.}}]{runge2019inferring}%
  \BibitemOpen
  \bibfield  {author} {\bibinfo {author} {\bibfnamefont {J.}~\bibnamefont
  {Runge}}, \bibinfo {author} {\bibfnamefont {S.}~\bibnamefont {Bathiany}},
  \bibinfo {author} {\bibfnamefont {E.}~\bibnamefont {Bollt}}, \bibinfo
  {author} {\bibfnamefont {G.}~\bibnamefont {Camps-Valls}}, \bibinfo {author}
  {\bibfnamefont {D.}~\bibnamefont {Coumou}}, \bibinfo {author} {\bibfnamefont
  {E.}~\bibnamefont {Deyle}}, \bibinfo {author} {\bibfnamefont
  {C.}~\bibnamefont {Glymour}}, \bibinfo {author} {\bibfnamefont
  {M.}~\bibnamefont {Kretschmer}}, \bibinfo {author} {\bibfnamefont {M.~D.}\
  \bibnamefont {Mahecha}}, \bibinfo {author} {\bibfnamefont {J.}~\bibnamefont
  {Mu{\~n}oz-Mar{\'\i}}},  \emph {et~al.},\ }\href@noop {} {\bibfield
  {journal} {\bibinfo  {journal} {Nature Communications}\ }\textbf {\bibinfo
  {volume} {10}},\ \bibinfo {pages} {2553} (\bibinfo {year}
  {2019})}\BibitemShut {NoStop}%
\bibitem [{\citenamefont {Dosi}\ and\ \citenamefont
  {Roventini}(2019)}]{Dosi:2019}%
  \BibitemOpen
  \bibfield  {author} {\bibinfo {author} {\bibfnamefont {G.}~\bibnamefont
  {Dosi}}\ and\ \bibinfo {author} {\bibfnamefont {A.}~\bibnamefont
  {Roventini}},\ }\href@noop {} {\bibfield  {journal} {\bibinfo  {journal} {J
  Evol Econ}\ }\textbf {\bibinfo {volume} {29}} (\bibinfo {year}
  {2019})}\BibitemShut {NoStop}%
\bibitem [{\citenamefont {Bressler}\ and\ \citenamefont
  {Seth}(2011)}]{bressler2011wiener}%
  \BibitemOpen
  \bibfield  {author} {\bibinfo {author} {\bibfnamefont {S.~L.}\ \bibnamefont
  {Bressler}}\ and\ \bibinfo {author} {\bibfnamefont {A.~K.}\ \bibnamefont
  {Seth}},\ }\href@noop {} {\bibfield  {journal} {\bibinfo  {journal}
  {Neuroimage}\ }\textbf {\bibinfo {volume} {58}},\ \bibinfo {pages} {323}
  (\bibinfo {year} {2011})}\BibitemShut {NoStop}%
\bibitem [{\citenamefont {Tononi}\ \emph {et~al.}(1994)\citenamefont {Tononi},
  \citenamefont {Sporns},\ and\ \citenamefont {Edelman}}]{Tononi1994}%
  \BibitemOpen
  \bibfield  {author} {\bibinfo {author} {\bibfnamefont {G.}~\bibnamefont
  {Tononi}}, \bibinfo {author} {\bibfnamefont {O.}~\bibnamefont {Sporns}}, \
  and\ \bibinfo {author} {\bibfnamefont {G.}~\bibnamefont {Edelman}},\ }\href
  {\doibase 10.1073/pnas.91.11.5033} {\bibfield  {journal} {\bibinfo  {journal}
  {Proceedings of the National Academy of Sciences}\ }\textbf {\bibinfo
  {volume} {91}},\ \bibinfo {pages} {5033} (\bibinfo {year}
  {1994})}\BibitemShut {NoStop}%
\bibitem [{\citenamefont {Balduzzi}\ and\ \citenamefont
  {Tononi}(2008)}]{Balduzzi2008}%
  \BibitemOpen
  \bibfield  {author} {\bibinfo {author} {\bibfnamefont {D.}~\bibnamefont
  {Balduzzi}}\ and\ \bibinfo {author} {\bibfnamefont {G.}~\bibnamefont
  {Tononi}},\ }\href {\doibase 10.1371/journal.pcbi.1000091} {\bibfield
  {journal} {\bibinfo  {journal} {PLoS Computational Biology}\ }\textbf
  {\bibinfo {volume} {4}},\ \bibinfo {pages} {1} (\bibinfo {year}
  {2008})}\BibitemShut {NoStop}%
\bibitem [{\citenamefont {Oizumi}\ \emph {et~al.}(2014)\citenamefont {Oizumi},
  \citenamefont {Albantakis},\ and\ \citenamefont {Tononi}}]{Oizumi2014}%
  \BibitemOpen
  \bibfield  {author} {\bibinfo {author} {\bibfnamefont {M.}~\bibnamefont
  {Oizumi}}, \bibinfo {author} {\bibfnamefont {L.}~\bibnamefont {Albantakis}},
  \ and\ \bibinfo {author} {\bibfnamefont {G.}~\bibnamefont {Tononi}},\ }\href
  {\doibase 10.1371/journal.pcbi.1003588} {\bibfield  {journal} {\bibinfo
  {journal} {PLoS Computational Biology}\ }\textbf {\bibinfo {volume} {10}}
  (\bibinfo {year} {2014}),\ 10.1371/journal.pcbi.1003588}\BibitemShut
  {NoStop}%
\bibitem [{\citenamefont {Seth}\ \emph {et~al.}(2011)\citenamefont {Seth},
  \citenamefont {Barrett},\ and\ \citenamefont {Barnett}}]{Seth2011}%
  \BibitemOpen
  \bibfield  {author} {\bibinfo {author} {\bibfnamefont {A.~K.}\ \bibnamefont
  {Seth}}, \bibinfo {author} {\bibfnamefont {A.~B.}\ \bibnamefont {Barrett}}, \
  and\ \bibinfo {author} {\bibfnamefont {L.}~\bibnamefont {Barnett}},\
  }\href@noop {} {\bibfield  {journal} {\bibinfo  {journal} {Philosophical
  Transactions of the Royal Society A: Mathematical, Physical and Engineering
  Sciences}\ }\textbf {\bibinfo {volume} {369}},\ \bibinfo {pages} {3748}
  (\bibinfo {year} {2011})}\BibitemShut {NoStop}%
\bibitem [{\citenamefont {van Walsum}\ \emph {et~al.}(2003)\citenamefont {van
  Walsum}, \citenamefont {Pijnenburg}, \citenamefont {Berendse}, \citenamefont
  {van Dijk}, \citenamefont {Knol}, \citenamefont {Scheltens},\ and\
  \citenamefont {Stam}}]{van2003neural}%
  \BibitemOpen
  \bibfield  {author} {\bibinfo {author} {\bibfnamefont {A.-M. v.~C.}\
  \bibnamefont {van Walsum}}, \bibinfo {author} {\bibfnamefont
  {Y.}~\bibnamefont {Pijnenburg}}, \bibinfo {author} {\bibfnamefont
  {H.}~\bibnamefont {Berendse}}, \bibinfo {author} {\bibfnamefont
  {B.}~\bibnamefont {van Dijk}}, \bibinfo {author} {\bibfnamefont
  {D.}~\bibnamefont {Knol}}, \bibinfo {author} {\bibfnamefont {P.}~\bibnamefont
  {Scheltens}}, \ and\ \bibinfo {author} {\bibfnamefont {C.}~\bibnamefont
  {Stam}},\ }\href@noop {} {\bibfield  {journal} {\bibinfo  {journal} {Clinical
  Neurophysiology}\ }\textbf {\bibinfo {volume} {114}},\ \bibinfo {pages}
  {1034} (\bibinfo {year} {2003})}\BibitemShut {NoStop}%
\bibitem [{\citenamefont {Mediano}\ \emph {et~al.}(2019)\citenamefont
  {Mediano}, \citenamefont {Seth},\ and\ \citenamefont
  {Barrett}}]{Mediano2019}%
  \BibitemOpen
  \bibfield  {author} {\bibinfo {author} {\bibfnamefont {P.}~\bibnamefont
  {Mediano}}, \bibinfo {author} {\bibfnamefont {A.}~\bibnamefont {Seth}}, \
  and\ \bibinfo {author} {\bibfnamefont {A.}~\bibnamefont {Barrett}},\
  }\href@noop {} {\bibfield  {journal} {\bibinfo  {journal} {Entropy}\ }\textbf
  {\bibinfo {volume} {21}},\ \bibinfo {pages} {17} (\bibinfo {year}
  {2019})}\BibitemShut {NoStop}%
\bibitem [{\citenamefont {Rosas}\ \emph {et~al.}(2019)\citenamefont {Rosas},
  \citenamefont {Mediano}, \citenamefont {Gastpar},\ and\ \citenamefont
  {Jensen}}]{Rosas2019}%
  \BibitemOpen
  \bibfield  {author} {\bibinfo {author} {\bibfnamefont {F.}~\bibnamefont
  {Rosas}}, \bibinfo {author} {\bibfnamefont {P.~A.~M.}\ \bibnamefont
  {Mediano}}, \bibinfo {author} {\bibfnamefont {M.}~\bibnamefont {Gastpar}}, \
  and\ \bibinfo {author} {\bibfnamefont {H.~J.}\ \bibnamefont {Jensen}},\
  }\href@noop {} {\enquote {\bibinfo {title} {Quantifying high-order
  interdependencies via multivariate extensions of the mutual information},}\ }
  (\bibinfo {year} {2019}),\ \Eprint {http://arxiv.org/abs/1902.11239}
  {arXiv:1902.11239 [cs.IT]} \BibitemShut {NoStop}%
\bibitem [{\citenamefont {James}\ \emph {et~al.}(2016)\citenamefont {James},
  \citenamefont {Barnett},\ and\ \citenamefont
  {Crutchfield}}]{james2016information}%
  \BibitemOpen
  \bibfield  {author} {\bibinfo {author} {\bibfnamefont {R.~G.}\ \bibnamefont
  {James}}, \bibinfo {author} {\bibfnamefont {N.}~\bibnamefont {Barnett}}, \
  and\ \bibinfo {author} {\bibfnamefont {J.~P.}\ \bibnamefont {Crutchfield}},\
  }\href@noop {} {\bibfield  {journal} {\bibinfo  {journal} {Physical Review
  Letters}\ }\textbf {\bibinfo {volume} {116}},\ \bibinfo {pages} {238701}
  (\bibinfo {year} {2016})}\BibitemShut {NoStop}%
\bibitem [{\citenamefont {Lizier}(2010)}]{Lizier2010}%
  \BibitemOpen
  \bibfield  {author} {\bibinfo {author} {\bibfnamefont {J.}~\bibnamefont
  {Lizier}},\ }\emph {\bibinfo {title} {The local information dynamics of
  distributed computation in complex systems}},\ \href@noop {} {Ph.D. thesis},\
  \bibinfo  {school} {University of Sydney} (\bibinfo {year}
  {2010})\BibitemShut {NoStop}%
\bibitem [{\citenamefont {Williams}\ and\ \citenamefont
  {Beer}(2010)}]{Williams2010}%
  \BibitemOpen
  \bibfield  {author} {\bibinfo {author} {\bibfnamefont {P.~L.}\ \bibnamefont
  {Williams}}\ and\ \bibinfo {author} {\bibfnamefont {R.~D.}\ \bibnamefont
  {Beer}},\ }\href@noop {} {\enquote {\bibinfo {title} {Nonnegative
  decomposition of multivariate information},}\ } (\bibinfo {year} {2010}),\
  \Eprint {http://arxiv.org/abs/1004.2515} {arXiv:1004.2515 [cs.IT]}
  \BibitemShut {NoStop}%
\bibitem [{\citenamefont {Crutchfield}\ and\ \citenamefont
  {Feldman}(2003)}]{crutchfield2003regularities}%
  \BibitemOpen
  \bibfield  {author} {\bibinfo {author} {\bibfnamefont {J.~P.}\ \bibnamefont
  {Crutchfield}}\ and\ \bibinfo {author} {\bibfnamefont {D.~P.}\ \bibnamefont
  {Feldman}},\ }\href@noop {} {\bibfield  {journal} {\bibinfo  {journal}
  {Chaos: An Interdisciplinary Journal of Nonlinear Science}\ }\textbf
  {\bibinfo {volume} {13}},\ \bibinfo {pages} {25} (\bibinfo {year}
  {2003})}\BibitemShut {NoStop}%
\bibitem [{\citenamefont {Grassberger}(1986)}]{grassberger1986toward}%
  \BibitemOpen
  \bibfield  {author} {\bibinfo {author} {\bibfnamefont {P.}~\bibnamefont
  {Grassberger}},\ }\href@noop {} {\bibfield  {journal} {\bibinfo  {journal}
  {International Journal of Theoretical Physics}\ }\textbf {\bibinfo {volume}
  {25}},\ \bibinfo {pages} {907} (\bibinfo {year} {1986})}\BibitemShut
  {NoStop}%
\bibitem [{\citenamefont {Crutchfield}\ \emph {et~al.}(2009)\citenamefont
  {Crutchfield}, \citenamefont {Ellison},\ and\ \citenamefont
  {Mahoney}}]{crutchfield2009time}%
  \BibitemOpen
  \bibfield  {author} {\bibinfo {author} {\bibfnamefont {J.~P.}\ \bibnamefont
  {Crutchfield}}, \bibinfo {author} {\bibfnamefont {C.~J.}\ \bibnamefont
  {Ellison}}, \ and\ \bibinfo {author} {\bibfnamefont {J.~R.}\ \bibnamefont
  {Mahoney}},\ }\href@noop {} {\bibfield  {journal} {\bibinfo  {journal}
  {Physical Review Letters}\ }\textbf {\bibinfo {volume} {103}},\ \bibinfo
  {pages} {094101} (\bibinfo {year} {2009})}\BibitemShut {NoStop}%
\bibitem [{Note1()}]{Note1}%
  \BibitemOpen
  \bibinfo {note} {We use $Y_1Y_2$ as a shorthand notation for the random
  vector $(Y_1,Y_2)$.}\BibitemShut {Stop}%
\bibitem [{Note2()}]{Note2}%
  \BibitemOpen
  \bibinfo {note} {$\protect \texttt {Syn}(X_1,X_2;Y_1Y_2)$ was in fact
  proposed in Ref. \cite {Griffith2014} as a measure of integrated
  information.}\BibitemShut {Stop}%
\bibitem [{Note3()}]{Note3}%
  \BibitemOpen
  \bibinfo {note} {In a general $N$-variable case, $\protect \mathcal {A}$ is
  the set of antichains on the lattice $(\protect \mathcal {P}(\protect \{1,
  ... , N\protect \}), \subseteq )$, discussed in \cite {Williams2010}. We
  focus on the bivariate case for clarity, although our results hold for any
  $N$.}\BibitemShut {Stop}%
\bibitem [{Note4()}]{Note4}%
  \BibitemOpen
  \bibinfo {note} {A proof of this is provided in the Appendix.}\BibitemShut
  {Stop}%
\bibitem [{Note5()}]{Note5}%
  \BibitemOpen
  \bibinfo {note} {We use the shorthand notation $\protect \bm {X}^{\alpha } =
  (X_{i_1},\protect \dots ,X_{i_K})$ for $\alpha = \protect \{i_1,\protect
  \dots ,i_K\protect \}$.}\BibitemShut {Stop}%
\bibitem [{Note6()}]{Note6}%
  \BibitemOpen
  \bibinfo {note} {Note that our framework does not prescribe a particular
  formula for $\protect \ensuremath {I_{\partial }^{\protect \{1\protect
  \}\protect \{2\protect \}\rightarrow \protect \{1\protect \}\protect
  \{2\protect \}}}\protect \xspace $. A discussion on this issue can be found
  in the supplementary material.}\BibitemShut {Stop}%
\bibitem [{Note7()}]{Note7}%
  \BibitemOpen
  \bibinfo {note} {The relation between $\Phi $ID and causal emergence~\cite
  {seth2010measuring} will be described in a separate publication.}\BibitemShut
  {Stop}%
\bibitem [{\citenamefont {Barrett}\ and\ \citenamefont
  {Seth}(2011)}]{Barrett2011}%
  \BibitemOpen
  \bibfield  {author} {\bibinfo {author} {\bibfnamefont {A.~B.}\ \bibnamefont
  {Barrett}}\ and\ \bibinfo {author} {\bibfnamefont {A.~K.}\ \bibnamefont
  {Seth}},\ }\href {\doibase 10.1371/journal.pcbi.1001052} {\bibfield
  {journal} {\bibinfo  {journal} {PLoS Computational Biology}\ }\textbf
  {\bibinfo {volume} {7}},\ \bibinfo {pages} {1} (\bibinfo {year}
  {2011})}\BibitemShut {NoStop}%
\bibitem [{\citenamefont {Griffith}(2014)}]{Griffith2014}%
  \BibitemOpen
  \bibfield  {author} {\bibinfo {author} {\bibfnamefont {V.}~\bibnamefont
  {Griffith}},\ }\href@noop {} {\enquote {\bibinfo {title} {A principled
  infotheoretic {$\varphi$}-like measure},}\ } (\bibinfo {year} {2014}),\
  \Eprint {http://arxiv.org/abs/1401.0978} {arXiv:1401.0978 [cs.IT]}
  \BibitemShut {NoStop}%
\bibitem [{\citenamefont {Oizumi}\ \emph {et~al.}(2016)\citenamefont {Oizumi},
  \citenamefont {Tsuchiya},\ and\ \citenamefont {Amari}}]{Oizumi2016}%
  \BibitemOpen
  \bibfield  {author} {\bibinfo {author} {\bibfnamefont {M.}~\bibnamefont
  {Oizumi}}, \bibinfo {author} {\bibfnamefont {N.}~\bibnamefont {Tsuchiya}}, \
  and\ \bibinfo {author} {\bibfnamefont {S.-i.}\ \bibnamefont {Amari}},\ }\href
  {\doibase 10.1073/pnas.1603583113} {\bibfield  {journal} {\bibinfo  {journal}
  {Proceedings of the National Academy of Sciences}\ }\textbf {\bibinfo
  {volume} {113}},\ \bibinfo {pages} {14817} (\bibinfo {year}
  {2016})}\BibitemShut {NoStop}%
\bibitem [{\citenamefont {James}\ \emph {et~al.}(2018)\citenamefont {James},
  \citenamefont {Emenheiser},\ and\ \citenamefont {Crutchfield}}]{James2018}%
  \BibitemOpen
  \bibfield  {author} {\bibinfo {author} {\bibfnamefont {R.~G.}\ \bibnamefont
  {James}}, \bibinfo {author} {\bibfnamefont {J.}~\bibnamefont {Emenheiser}}, \
  and\ \bibinfo {author} {\bibfnamefont {J.~P.}\ \bibnamefont {Crutchfield}},\
  }\href@noop {} {\bibfield  {journal} {\bibinfo  {journal} {Journal of Physics
  A: Mathematical and Theoretical}\ }\textbf {\bibinfo {volume} {52}},\
  \bibinfo {pages} {014002} (\bibinfo {year} {2018})}\BibitemShut {NoStop}%
\bibitem [{Note999()}]{Note999}%
  \BibitemOpen
  \bibinfo {note} {Note that the original definition of causal density is
  normalised by $L(L-1)$, and has been proven to be bounded by mutual
  information \cite {Mediano2019}.}\BibitemShut {Stop}%
\bibitem [{\citenamefont {Williams}\ and\ \citenamefont
  {Beer}(2011)}]{Williams2011}%
  \BibitemOpen
  \bibfield  {author} {\bibinfo {author} {\bibfnamefont {P.~L.}\ \bibnamefont
  {Williams}}\ and\ \bibinfo {author} {\bibfnamefont {R.~D.}\ \bibnamefont
  {Beer}},\ }\href@noop {} {\bibfield  {journal} {\bibinfo  {journal} {arXiv
  preprint arXiv:1102.1507}\ } (\bibinfo {year} {2011})}\BibitemShut {NoStop}%
\bibitem [{\citenamefont {Feldman}\ and\ \citenamefont
  {Crutchfield}(1998)}]{Feldman1998}%
  \BibitemOpen
  \bibfield  {author} {\bibinfo {author} {\bibfnamefont {D.~P.}\ \bibnamefont
  {Feldman}}\ and\ \bibinfo {author} {\bibfnamefont {J.~P.}\ \bibnamefont
  {Crutchfield}},\ }\href@noop {} {\bibfield  {journal} {\bibinfo  {journal}
  {Physics Letters A}\ }\textbf {\bibinfo {volume} {238}},\ \bibinfo {pages}
  {244} (\bibinfo {year} {1998})}\BibitemShut {NoStop}%
\bibitem [{Note8()}]{Note8}%
  \BibitemOpen
  \bibinfo {note} {For example, measures that accurately discriminate between
  neural configurations corresponding to conscious and unconscious states in a
  particular experimental paradigm~\cite {Casali2013}.}\BibitemShut {Stop}%
\bibitem [{Note998()}]{Note998}%
  \BibitemOpen
  \bibinfo {note} {Intuitively, a causal analysis reveals what the system
  \protect \emph {could do}, while a dynamical analysis based on attractor
  statistics reveals what the system \protect \emph {actually
  does}.}\BibitemShut {Stop}%
\bibitem [{\citenamefont {Pearl}\ and\ \citenamefont
  {Mackenzie}(2018)}]{pearl2018book}%
  \BibitemOpen
  \bibfield  {author} {\bibinfo {author} {\bibfnamefont {J.}~\bibnamefont
  {Pearl}}\ and\ \bibinfo {author} {\bibfnamefont {D.}~\bibnamefont
  {Mackenzie}},\ }\href@noop {} {\emph {\bibinfo {title} {The book of why:
  {The} new science of cause and effect}}}\ (\bibinfo  {publisher} {Basic
  Books},\ \bibinfo {year} {2018})\BibitemShut {NoStop}%
\bibitem [{Note9()}]{Note9}%
  \BibitemOpen
  \bibinfo {note} {Also, note that the maximum entropy distributions employed
  by some causal integration frameworks are well-defined for discrete Markovian
  systems, but in general may not always exist \cite
  {Barrett2019}.}\BibitemShut {Stop}%
\bibitem [{\citenamefont {Charalambides}(2002)}]{charalambides2002enumerative}%
  \BibitemOpen
  \bibfield  {author} {\bibinfo {author} {\bibfnamefont {C.~A.}\ \bibnamefont
  {Charalambides}},\ }\href@noop {} {\emph {\bibinfo {title} {Enumerative
  combinatorics}}}\ (\bibinfo  {publisher} {Chapman and Hall/CRC},\ \bibinfo
  {year} {2002})\BibitemShut {NoStop}%
\bibitem [{\citenamefont {Crampton}\ and\ \citenamefont
  {Loizou}(2001)}]{Crampton2001}%
  \BibitemOpen
  \bibfield  {author} {\bibinfo {author} {\bibfnamefont {J.}~\bibnamefont
  {Crampton}}\ and\ \bibinfo {author} {\bibfnamefont {G.}~\bibnamefont
  {Loizou}},\ }\href@noop {} {\bibfield  {journal} {\bibinfo  {journal}
  {International Mathematical Journal}\ }\textbf {\bibinfo {volume} {1}},\
  \bibinfo {pages} {223} (\bibinfo {year} {2001})}\BibitemShut {NoStop}%
\bibitem [{\citenamefont {Quax}\ \emph {et~al.}(2017)\citenamefont {Quax},
  \citenamefont {Har-Shemesh},\ and\ \citenamefont
  {Sloot}}]{quax2017quantifying}%
  \BibitemOpen
  \bibfield  {author} {\bibinfo {author} {\bibfnamefont {R.}~\bibnamefont
  {Quax}}, \bibinfo {author} {\bibfnamefont {O.}~\bibnamefont {Har-Shemesh}}, \
  and\ \bibinfo {author} {\bibfnamefont {P.}~\bibnamefont {Sloot}},\
  }\href@noop {} {\bibfield  {journal} {\bibinfo  {journal} {Entropy}\ }\textbf
  {\bibinfo {volume} {19}},\ \bibinfo {pages} {85} (\bibinfo {year}
  {2017})}\BibitemShut {NoStop}%
\bibitem [{\citenamefont {Rassouli}\ \emph {et~al.}(2018)\citenamefont
  {Rassouli}, \citenamefont {Rosas},\ and\ \citenamefont
  {G{\"u}nd{\"u}z}}]{rassouli2018latent}%
  \BibitemOpen
  \bibfield  {author} {\bibinfo {author} {\bibfnamefont {B.}~\bibnamefont
  {Rassouli}}, \bibinfo {author} {\bibfnamefont {F.}~\bibnamefont {Rosas}}, \
  and\ \bibinfo {author} {\bibfnamefont {D.}~\bibnamefont {G{\"u}nd{\"u}z}},\
  }in\ \href@noop {} {\emph {\bibinfo {booktitle} {2018 IEEE International
  Workshop on Information Forensics and Security (WIFS)}}}\ (\bibinfo
  {organization} {IEEE},\ \bibinfo {year} {2018})\ pp.\ \bibinfo {pages}
  {1--7}\BibitemShut {NoStop}%
\bibitem [{\citenamefont {Rosas}\ \emph {et~al.}(2016)\citenamefont {Rosas},
  \citenamefont {Ntranos}, \citenamefont {Ellison}, \citenamefont {Pollin},\
  and\ \citenamefont {Verhelst}}]{rosas2016understanding}%
  \BibitemOpen
  \bibfield  {author} {\bibinfo {author} {\bibfnamefont {F.}~\bibnamefont
  {Rosas}}, \bibinfo {author} {\bibfnamefont {V.}~\bibnamefont {Ntranos}},
  \bibinfo {author} {\bibfnamefont {C.}~\bibnamefont {Ellison}}, \bibinfo
  {author} {\bibfnamefont {S.}~\bibnamefont {Pollin}}, \ and\ \bibinfo {author}
  {\bibfnamefont {M.}~\bibnamefont {Verhelst}},\ }\href@noop {} {\bibfield
  {journal} {\bibinfo  {journal} {Entropy}\ }\textbf {\bibinfo {volume} {18}},\
  \bibinfo {pages} {38} (\bibinfo {year} {2016})}\BibitemShut {NoStop}%
\bibitem [{\citenamefont {Seth}(2010)}]{seth2010measuring}%
  \BibitemOpen
  \bibfield  {author} {\bibinfo {author} {\bibfnamefont {A.~K.}\ \bibnamefont
  {Seth}},\ }\href@noop {} {\bibfield  {journal} {\bibinfo  {journal}
  {Artificial life}\ }\textbf {\bibinfo {volume} {16}},\ \bibinfo {pages} {179}
  (\bibinfo {year} {2010})}\BibitemShut {NoStop}%
\bibitem [{\citenamefont {Casali}\ \emph {et~al.}(2013)\citenamefont {Casali},
  \citenamefont {Gosseries}, \citenamefont {Rosanova}, \citenamefont {Boly},
  \citenamefont {Sarasso}, \citenamefont {Casali}, \citenamefont {Casarotto},
  \citenamefont {Bruno}, \citenamefont {Laureys}, \citenamefont {Tononi},\ and\
  \citenamefont {Massimini}}]{Casali2013}%
  \BibitemOpen
  \bibfield  {author} {\bibinfo {author} {\bibfnamefont {A.~G.}\ \bibnamefont
  {Casali}}, \bibinfo {author} {\bibfnamefont {O.}~\bibnamefont {Gosseries}},
  \bibinfo {author} {\bibfnamefont {M.}~\bibnamefont {Rosanova}}, \bibinfo
  {author} {\bibfnamefont {M.}~\bibnamefont {Boly}}, \bibinfo {author}
  {\bibfnamefont {S.}~\bibnamefont {Sarasso}}, \bibinfo {author} {\bibfnamefont
  {K.~R.}\ \bibnamefont {Casali}}, \bibinfo {author} {\bibfnamefont
  {S.}~\bibnamefont {Casarotto}}, \bibinfo {author} {\bibfnamefont {M.-A.}\
  \bibnamefont {Bruno}}, \bibinfo {author} {\bibfnamefont {S.}~\bibnamefont
  {Laureys}}, \bibinfo {author} {\bibfnamefont {G.}~\bibnamefont {Tononi}}, \
  and\ \bibinfo {author} {\bibfnamefont {M.}~\bibnamefont {Massimini}},\ }\href
  {\doibase 10.1126/scitranslmed.3006294} {\bibfield  {journal} {\bibinfo
  {journal} {Science Translational Medicine}\ }\textbf {\bibinfo {volume}
  {5}},\ \bibinfo {pages} {198ra105} (\bibinfo {year} {2013})}\BibitemShut
  {NoStop}%
\bibitem [{\citenamefont {Barrett}\ and\ \citenamefont
  {Mediano}(2019)}]{Barrett2019}%
  \BibitemOpen
  \bibfield  {author} {\bibinfo {author} {\bibfnamefont {A.}~\bibnamefont
  {Barrett}}\ and\ \bibinfo {author} {\bibfnamefont {P.}~\bibnamefont
  {Mediano}},\ }\href
  {https://www.ingentaconnect.com/content/imp/jcs/2019/00000026/f0020001/art00002}
  {\bibfield  {journal} {\bibinfo  {journal} {Journal of Consciousness
  Studies}\ }\textbf {\bibinfo {volume} {26}},\ \bibinfo {pages} {11} (\bibinfo
  {year} {2019})}\BibitemShut {NoStop}%
\end{thebibliography}%

\end{document}